\documentclass[letter, traditabstract]{aa}  
\usepackage{graphicx}
\usepackage{txfonts}
\usepackage{natbib}
\bibpunct{(}{)}{;}{a}{}{,}	
\begin{document}
   \title{Connection between dense gas mass fraction, turbulence driving, and star formation efficiency of molecular clouds}


   \author{J. Kainulainen	\inst{1},
           C. Federrath 		\inst{2}, \and
           T. Henning		\inst{1}          }
   \offprints{jtkainul@mpia.de}

   \institute{Max-Planck-Institute for Astronomy, K\"onigstuhl 17, 69117 Heidelberg, Germany \\
              \email{jtkainul@mpia.de} \and
               Monash Centre for Astrophysics, School of Mathematical Sciences, Monash University, Vic 3800, Australia \\
            }
   \date{Received ; accepted }
\abstract{We examine the physical parameters that affect the accumulation of gas in molecular clouds to high column densities where the formation of stars takes place. In particular, we analyze the dense gas mass fraction (DGMF) in a set of self-gravitating, isothermal, magnetohydrodynamic turbulence simulations including sink particles to model star formation. We find that the simulations predict close to exponential DGMFs over the column density range $N(\mathrm{H}_2) = 3-25 \times 10^{21}$ cm$^{-2}$ that can be easily probed via, e.g., dust extinction measurements. The exponential slopes correlate with the type of turbulence driving and also with the star formation efficiency. They are almost uncorrelated with the sonic Mach number and magnetic-field strength. The slopes at early stages of cloud evolution are steeper than at the later stages.  A comparison of these predictions with observations shows that only simulations with relatively non-compressive driving ($b \lesssim 0.4$) agree with the DGMFs of nearby molecular clouds. Massive infrared dark clouds can show DGMFs that are in agreement with more compressive driving. The DGMFs of molecular clouds can be significantly affected by how compressive the turbulence is on average. Variations in the level of compression can cause scatter to the DGMF slopes, and some variation is indeed necessary to explain the spread of the observed DGMF slopes. The observed DGMF slopes can also be affected by the clouds' star formation activities and statistical cloud-to-cloud variations.
} 
  \authorrunning{J. Kainulainen et al.}
  \titlerunning{Dense gas mass fraction of molecular clouds}
 
  \maketitle


\section{Introduction} 
\label{sec:intro}


Star formation is ultimately controlled by the processes that regulate the formation of density enhancements in molecular clouds. In our current picture, the density statistics of the interstellar medium are heavily affected by supersonic turbulence \citep[for a review, see][]{hen12}. 
The density statistics depend on characteristics such as the total turbulent and magnetic energy \citep[e.g.,][FK13 hereafter]{pad97mnras, nor99, vaz01, kow07, mol12, fed13}, the driving mechanism of the turbulence \citep[e.g.,][FK12, hereafter]{fed10, fed12}, the equation of state \citep[e.g.,][]{pas98, gaz13}, and the driving scale \citep[e.g.,][]{fis04, bru09}. Constraining these characteristics is fundamental for virtually all analytic star formation theories.


We have previously employed near-infrared dust extinction mapping in analyzing column density statistics of molecular clouds \citep[][KT13 hereafter]{kai09, kai11a, kai11b, kai13}. This technique is sensitive and well-calibrated at low column densities, making it suitable to study the mass reservoirs of molecular clouds. Exploiting this advantage, we studied how the clouds gather gas to the regime where star formation occurs. We used an easily accessible characteristic to quantify this, namely the dense gas mass fraction\footnote{We purposefully use here the term "dense gas mass fraction" instead of "cumulative mass function" (CMF) from our previous works. This is to avoid confusion with the "core mass function" that is commonly used in literature.} (DGMF, hereafter), defined as a function that gives the fraction of the cloud's mass above a column density value
\begin{equation}
\mathrm{d}M' (> N) = \frac{M(> N)}{M_\mathrm{tot}},
\label{eq:dgmf}
\end{equation}
where $M(> N)$ is the mass above the column density $N$ and $M_\mathrm{tot}$ is the total mass. The DGMF is linked to the probability density function (PDF), $p(N)$, of column densities, which gives the probability to have a column density between $[N, N+dN]$, via
\begin{equation}
dM' = \int_N^{N_\mathrm{high}} p(N') dN' / \int_{N_\mathrm{low}}^{N_\mathrm{high}} p(N') dN',
\label{eq:dgmf-pdf}
\end{equation}
where $[N_\mathrm{low}, N_\mathrm{high}]$ is the probed column density range. The reason for analyzing DGMFs instead of PDFs is simply the intuitive connection to the total mass reservoir of the cloud. Previously, DGMFs have been analyzed by, e.g., \citet{kai09} who showed that starless clouds contain much less dense gas than star-forming clouds and by \citet{lad10} who used them to derive a star-formation threshold.

From the theoretical point-of-view, the form of the DGMF can be controlled by any of the forces affecting the cloud's density structure. The key parameters describing these forces are\footnote{However, see the discussion on the caveat related to the Reynolds numbers of simulations in Section \ref{subsec:sim-DGMFs}} \emph{i)} the sonic Mach number, ${\cal M}_\mathrm{s}$, \emph{ii)} the turbulence \emph{driving} \citep{fed08, fed10}, which is commonly denoted by $b$, with $b=1/3$ corresponding to purely solenoidal driving and $b=1$ to fully compressive driving, and \emph{iii)} the magnetic field strength, $B$, reflected by the Alfv\'en Mach number, ${\cal M}_\mathrm{A}$. These parameters relate to density fluctuations via \citep[][]{nor99, pri11, mol12}
\begin{equation}
\sigma^2_{\ln{\rho / \langle \rho \rangle}} = \ln{(1+b^2 {\cal M_\mathrm{s}}^2   \frac{\beta}{\beta+1})},
\label{eq:b}
\end{equation}
where $\sigma_{\ln{\rho / \langle \rho \rangle}}$ is the standard deviation of logarithmic, mean-normalized densities and $\beta = 2{\cal M}^2_\mathrm{A} / {\cal M}^2_\mathrm{s}$. This form of Eq. \ref{eq:b} \citep{mol12} is valid up to moderate magnetic field strengths, ${\cal M}_\mathrm{A} \gtrsim 6$. The strength of the ${\cal M}_\mathrm{s}$ - density coupling is of great importance for analytic star formation theories, because it directly affects the star formation rates and -efficiencies (SFE) they predict \citep[e.g.,][see FK12]{kru05, hen11, pad11}. 

In this work, we will estimate how the different physical parameters affect the observed DGMFs of molecular clouds. To this goal, we will analyze numerical turbulence simulations and derive predictions for observable DGMFs. We will then compare the predictions to the results of \citet{kai09, kai11b} and KT13 \citep[see also][]{lad10}.

\section{Simulation data}           
\label{sec:data}



We analyze a set of magneto-hydrodynamic simulations of isothermal, driven turbulence in a periodic box, including self-gravity and sink particles to follow gas accretion onto protostars (see FK12). Each simulation is a time-series which starts ($t=0$) when the turbulence is fully developed and the gravity is switched on. Then, the evolution is followed as a function of SFE, defined as the fraction of mass accreted into sink particles. The formation of the first sink particle occurs at $\mathrm{SFE=0\%}$.
%
%
The sink particles affect their surroundings because of gas accretion, and we eliminated them from the simulations. The issue is described in Appendix \ref{app:sinks}. Here we quote the main result: the DGMFs of ${\cal M}_\mathrm{s} = 10$ simulations (which we directly compare with observations) with $512^3$ cells are unaffected by sink particles below $N(\mathrm{H}_2) < 11 \times 10^{21}$ cm$^{-2}$. They are $70$\% accurate up to $N(\mathrm{H}_2) \approx 25 \times 10^{21}$ cm$^{-2}$. We also show in Appendix \ref{app:resolution} that the resolution does not affect the DGMFs in this range.


   \begin{figure*}
   \centering
\includegraphics[bb = 0 4 960 225, clip=true, width=\textwidth]{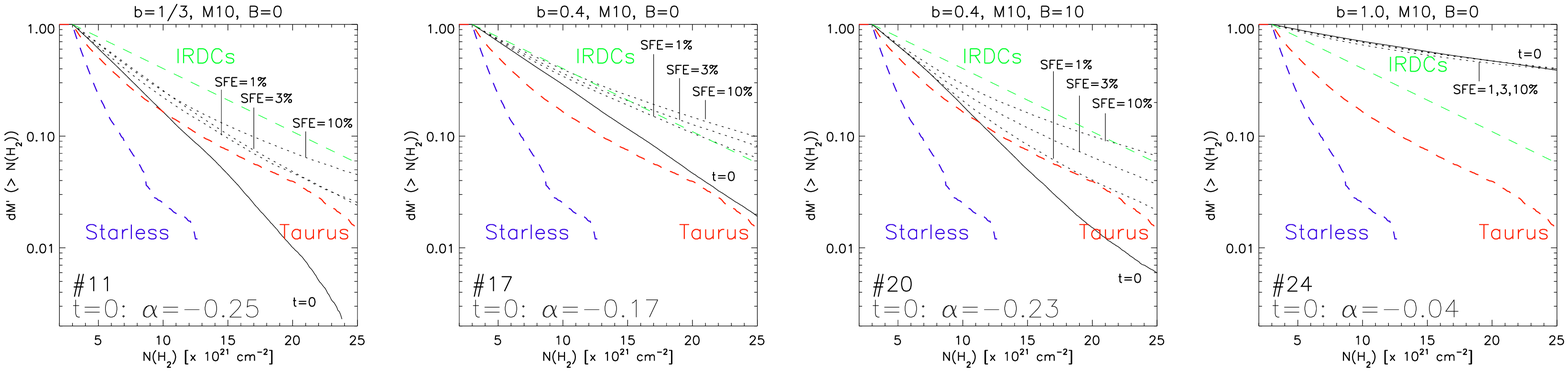}
      \caption{DGMFs of four simulations (black lines) with ${\cal M}_\mathrm{s} = 10$, processed to mimic those observed with near-infrared dust extinction mapping technique. The solid lines show the DGMFs at $t=0$ and the dotted lines at time-steps $\mathrm{SFE} = \{1, 3, 10\}\%$. The panels also show with dashed lines the mean DGMF of nearby starless clouds (blue) and of Taurus \citep[][red]{kai09}, and of a sample of IRDCs (KT13, green). 
                    }
         \label{fig:DGMFs-nir}
   \end{figure*}


The simulations were scaled so that their virial parameters, $\alpha_\mathrm{vir, 0} = 5 \sigma^2_\mathrm{v} L / (6 G M)$ where $\sigma_\mathrm{v}$ is the 3-D velocity dispersion and $L$ the size of the simulation, were close to unity. Observations have shown that molecular clouds, on average, show $\alpha_\mathrm{vir, 0} \approx 1$ \citep[e.g.,][]{hey09}. However, this definition is an idealized approximation. The actual virial parameters, $\alpha_\mathrm{vir} = 2|E_\mathrm{kin}|/|E_\mathrm{pot}|$, vary by more than an order of magnitude in the simulations. However, the actual virial parameters do not affect density PDFs greatly (FK12). If the virial parameter is "low-enough" to allow some collapse, the density structure is determined by other parameters \citep[FK12;][]{mol12}.

To make a realistic comparison with observations, we processed the simulations with ${\cal M}_\mathrm{s} = 5-10$ to mimic data derived using near-infrared dust extinction mapping \citep{lom01}. First, column density data from simulations was re-gridded to $60 \arcsec / \mathrm{pixel}$ and smoothed to 
the $FWHM = 120\arcsec$ resolution (0.09 pc at $150$ pc distance). The native resolution of the simulations with ${\cal M}_\mathrm{s} > 10$ is coarser than this, and we could not smooth them (we do not compare them with the lower ${\cal M}_\mathrm{s}$ simulations). Then, the column densities outside $N(\mathrm{H}_2) = [3, 25] \times 10^{21}$ cm$^{-2}$ were discarded, approximating the dynamic range of extinction mapping. The lower limit of the range was chosen to be high enough that it is possible to define separate structures in simulations using (approximately) closed contours of constant column density. This is because observationally, "clouds" are commonly defined in this manner \citep[e.g.,][]{lad10}. Finally, Gaussian noise with $\sigma (N) = 0.018N(\mathrm{H}_\mathrm{2}) + 0.2 \times 10^{21}$ cm$^{-2}$ was added, following typical uncertainties in the data of \citet{kai09}. This procedure was repeated for three different projections of the simulation data, and the DGMFs from them were averaged to form the final DGMF.

We examined the effects of the resolution and noise to the DGMFs. We experimented with the resolution of 0.03 pc that studies employing \emph{Herschel} data of nearby clouds will reach \citep[e.g.,][]{sch13}. Similar resolution is reached by combined near- and mid-infrared extinction mapping when applied to infrared dark clouds (IRDCs, KT13). The effect of the resolution and noise to the DGMFs was practically negligible.

\section{Results and Discussion}           
\label{sec:results}
\label{sec:discussion}

\subsection{Dependence of the DGMF on physical parameters}           
\label{subsec:sim-DGMFs}


We derived the DGMFs for the simulations up to $\mathrm{SFE} = 10$\%. Figure \ref{fig:DGMFs-nir} shows the DGMFs of four simulations with ${\cal M}_\mathrm{s}=10$ and $b = \{1/3, 0.4, 1\}$. For the case $b=0.4$, a non-magnetized and magnetized simulation is shown. The DGMFs at early stages ($t = 0$ and $\mathrm{SFE} = 0$\%) are well-described by exponential functions, $dM' \propto e^{\alpha N}$. When star formation begins, the DGMFs flatten. Their shapes remain close to an exponential function, or curve upwards approaching a powerlaw shape. This behavior is similar in all models. Since the DGMFs are close to exponential functions in the range $N(\mathrm{H_2}) = 3-11 \times 10^{21}$ cm$^{-2}$, we quantified their shapes through fits of exponentials. This yielded the range $\alpha = [-0.41, -0.023]$ in all models.

\begin{figure*}
   \centering
\includegraphics[bb= 20 5 480 355, clip=true, width=0.33\textwidth]{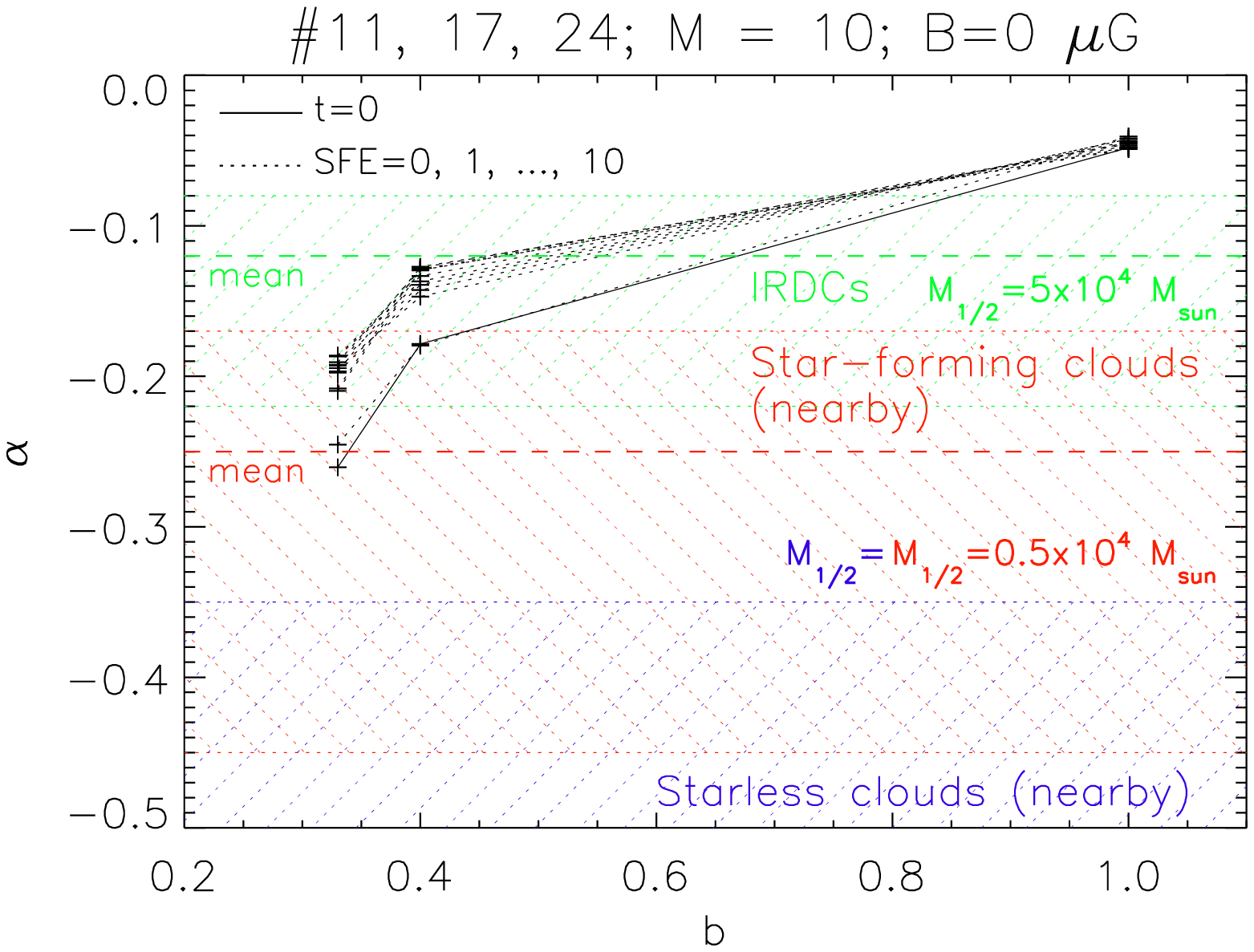}\includegraphics[bb= 20 5 480 355, clip=true, width=0.33\textwidth]{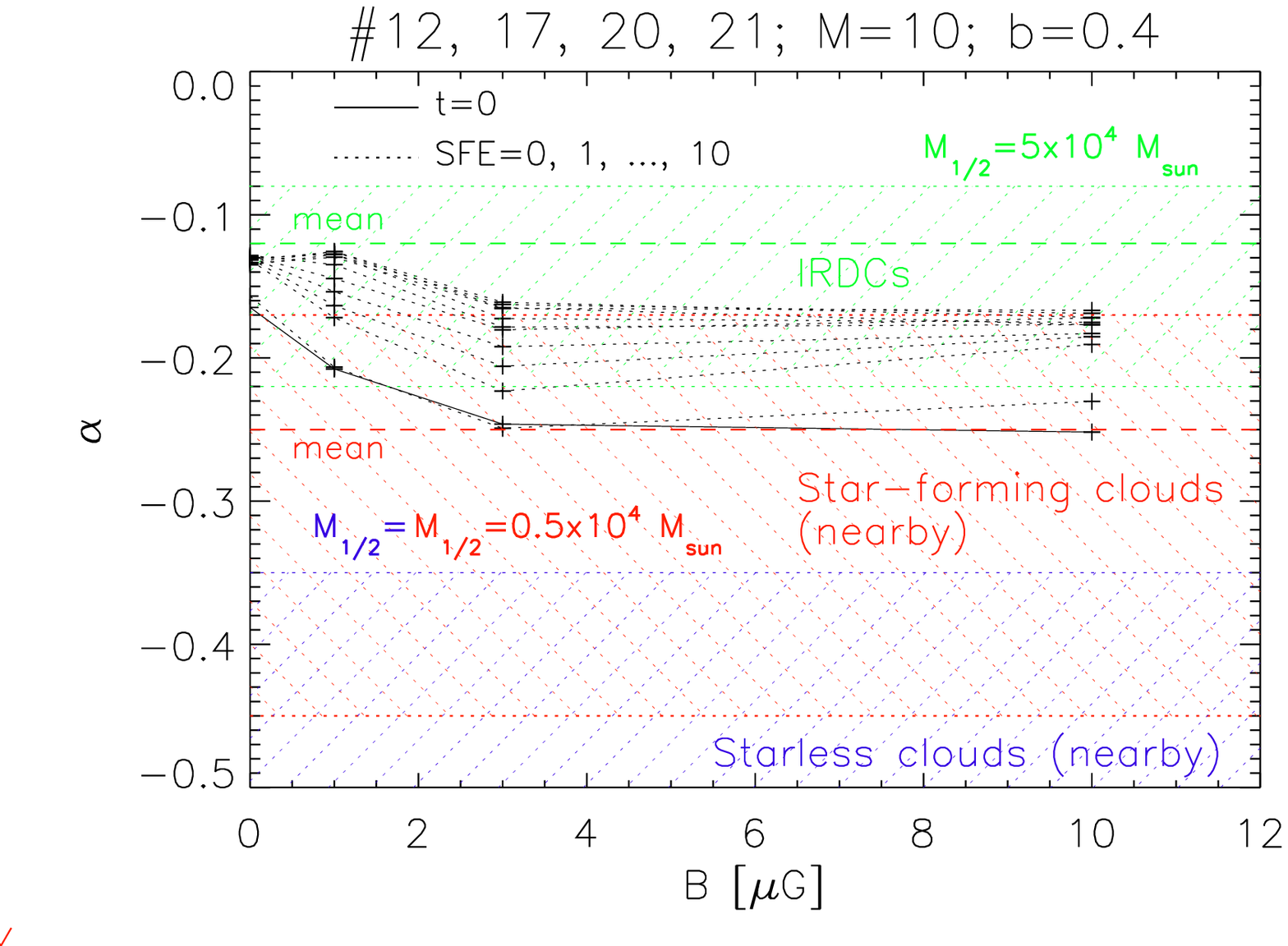}\includegraphics[bb= 20 5 480 355, clip=true, width=0.33\textwidth]{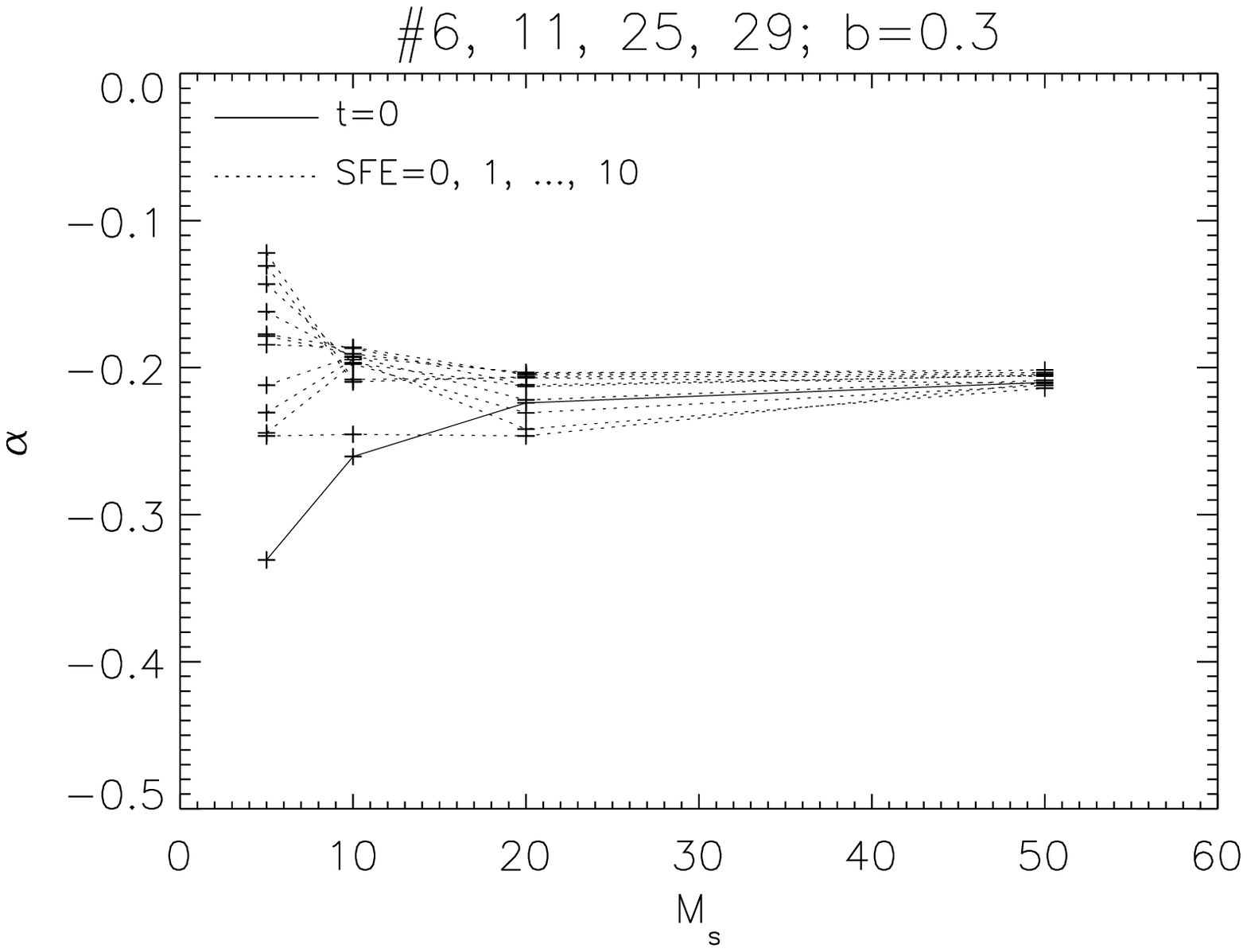}
      \caption{Exponential slopes of the DGMFs as a function of $b$ (\emph{left}), $B$ (\emph{center}), and ${\cal M}_\mathrm{s}$ (\emph{right}). The solid black lines show the timestep $t = 0$, and the dotted lines $\mathrm{SFE}=\{0, 1, \dots, 10\}\%$. The blue, red, and green shaded regions indicate the slopes observed in starless and star-forming nearby clouds \citep{kai09} and in IRDCs (KT13, however, see the discussion on these data in Section \ref{subsec:comp_with_obs}). The median masses of each set of the clouds, $M_\mathrm{1/2}$, are shown in the panels.
                    }
         \label{fig:slopes}
   \end{figure*}

We examined the dependence of the DGMF slopes on the driving of turbulence and magnetic field strength ($B$) in the simulations with ${\cal M}_\mathrm{s} = 10$. The results are shown in Fig. \ref{fig:slopes} (left and center). Most importantly, \emph{the DGMF slope responds most sensitively to the turbulence driving}, changing by a factor of $4.8-8.5$ when $b$ changes from $1/3$ to 1. The slopes depend clearly less on $B$. The non-magnetic simulations show significantly shallower slopes than magnetized ones, but if $ B \gtrsim 3$ $\mu$G, the slopes are uncorrelated with it. 


The DGMF slopes depend on the SFE. The dependency is stronger in magnetized than in non-magnetized simulations: the spreads of the slopes in the range $\mathrm{SFE}=[1, 10]\%$ for these cases are $0.09$ and 0.03, respectively. The mean difference in the slopes of non-magnetized and magnetized runs is 0.05. The early stages ($t=0$, $\mathrm{SFE}=0$\%) show clearly steeper slopes than the higher SFEs. 
%
%
We also examined the relationship between the DGMF slopes and ${\cal M}_\mathrm{s}$. For this, we derived the DGMFs in the native resolution of the simulations (smoothing would greatly reduce the size of the low-${\cal M}_\mathrm{s}$ runs). Therefore, the results should be compared to observations with caution. Figure \ref{fig:slopes} shows the DGMF slopes and ${\cal M}_\mathrm{s}$ in simulations with $b=1/3$. The slopes are non-responsive to ${\cal M}_\mathrm{s}$, except when ${\cal M}_\mathrm{s} = 5$. 


The DGMFs can vary also due to \emph{i)} the random nature of turbulence ("cloud-to-cloud" variations) and \emph{ii)} projection effects. The former can be examined by comparing simulations that have the same input parameters, but different random number seeds (e.g., \#12, 14, and 17, see Table \ref{tab:sinks}). Unfortunately, we only had three simulation pairs with varying random number seeds. The mean difference in the DGMF slopes among these was 0.08 at the early stages ($t = 0$, $\mathrm{SFE} = 0\%$). However, for time-steps $\mathrm{SFE} \ge 1$ the mean difference was only 0.02. The projection effects were studied by examining the standard deviation of the slopes derived for three different projections of all models. The mean standard deviation of the slopes in all models was 0.03.


We note that the effective Reynolds numbers of our simulations ($\lesssim$$10^4$) are lower than that of the interstellar medium ($\sim$$10^7$). It is not clear how this affects the predicted statistical properties. \citet{alu13} has rigorously shown that the direct influence of driving on the kinetic energy is restricted to scales larger than the smallest scale at which the turbulence is stirred. However, numerical \citep{fed10} and analytic \citep{gal11} works have found differences in flow statistics in the range that can be considered to be the "inertial range" of compressible turbulence simulations. Resolution studies of the simulations suggest that the driving-induced differences remain when the Reynolds number increases. As this issue cannot be addressed with the current computational methods, our results are also subject to it.

\subsection{Comparing the predictions with observations}           
\label{subsec:comp_with_obs}

Figures \ref{fig:DGMFs-nir} and \ref{fig:slopes} show observed DGMFs to be compared with the simulated ones. Figure \ref{fig:DGMFs-nir} shows the mean DGMF of quiescent clouds (LDN1719, Lupus V, Cha III, and Musca) and a DGMF of a typical star-forming cloud (Taurus) from \citet{kai09}, and a mean DGMF of ten IRDCs from KT13. Figure \ref{fig:slopes} shows the ranges of the observed slopes from \citet{kai09}, which span $\alpha = [-0.17, -0.45]$ for 13 nearby star-forming clouds and $\alpha = [-0.35, -1.2]$ for four quiescent clouds. The range of IRDC slopes from KT13 is also shown. We note that the DGMFs of IRDCs in KT13 were derived from a slightly different column density range than those of nearby clouds (they begin from $N(\mathrm{H}_2) \approx 7 \times 10^{21}$ cm$^{-2}$). Thus, the comparison of them with the other data should be considered only suggestive.

%
%

The dependence of the DGMF slopes on the turbulence driving allows us to constrain $b$ (see Fig. \ref{fig:slopes}). None of the simulations shows as steep slopes as observed in starless clouds. From the non-magnetized simulations, only those with $b=1/3$ are in agreement with the nearby star-forming clouds.  Magnetic fields can steepen the slopes by about $0.05$ (Fig. \ref{fig:slopes}, center). Therefore, from the magnetized runs those with $b=1/3$, or $b=0.4$ and $B \ge 3 \ \mu$G agree with star-forming clouds. \emph{The fully compressive simulations produce a greatly higher fraction of dense gas than observed in nearby clouds}. The comparison suggests a low $b$ for nearby molecular clouds on average, possibly lower than previously estimated by \citet{pad97apj} and \citet{bru10taurus} in Taurus, $b \approx 0.5$.

The DGMF slopes correlate with the SFE, depending on whether the cloud is magnetized or not. Since in the current view clouds have magnetic fields \citep[][]{cru12}, the spread of slopes is likely the most realistic in magnetized simulations (i.e., 0.1, see Fig. \ref{fig:slopes}). Thus, it seems that part of the spread in the observed slopes originates from the SFEs of the clouds. We used a Monte Carlo simulation to estimate whether all the variation in the observed slopes can originate from changes in the SFE and statistical variations. We assumed that the changes due to SFE are uniformly distributed between $[0, 0.1]$ and the statistical variations are normally distributed with $\sigma = 0.04$. The test showed that the probability that 13 clouds span a range $>0.28$ is $0.2\%$. Note that the range of the observed slopes can be wider. KT13 showed that IRDCs possibly have flatter DGMFs than nearby clouds (Fig. \ref{fig:slopes}). In conclusion, it seems likely that the spread of the observed DGMF slopes cannot be explained by statistical variations and changes in the SFE alone. Changes in the clouds' average compression provides one possible source to account for this variation.

One interesting question for the future is to examine the effect of cloud mass to the DGMFs. There are no very massive clouds in the nearby cloud sample (median mass $0.5 \times 10^4$ M$_\odot$). In contrast, the median mass of the IRDCs is $5 \times 10^4$ M$_\odot$, which is ten times higher. This could contribute to the differences seen in the slopes of the two cloud sets. However as discussed earlier, comparing DGMFs of IRDCs with nearby clouds is not without caveats. The question could be properly addressed by a study of a statistical sample of IRDCs, or a study of the nearest high-mass clouds (e.g., Orion, Cygnus, Rosette) employing \emph{Herschel} data.

The weak dependence of the DGMF slopes on ${\cal M}_\mathrm{s}$ appears to be an effect of the narrow column density range we examine (note that the results were derived from simulations that have differing physical resolutions and are only suggestive). The density PDF is expected to respond to ${\cal M}_\mathrm{s}$ following Eq. \ref{eq:b}, which should reflect to the DGMFs. However, it appears that in the range of $N(\mathrm{H}_2) = 3-11 \times 10^{21}$ cm$^{-2}$ the effect is insignificant. This result is in agreement with \citet{goo09} who did not detect any dependence between column density PDF widths and CO linewidths in Perseus. However, we recently measured the column density PDF widths using a high-dynamic-range technique (KT13) and concluded that if a wider range is examined, the PDF widths correlate with ${\cal M}_\mathrm{s}$. 


When comparing observed DGMFs with simulations, it should be kept in mind that in simulations "driving" is well-defined and ideal: energy is injected at large scales, with certain characteristics such as the divergence and curl. In real clouds, energy is likely injected at multiple scales and the characteristics of the driving can depend on the scale. However, if some of these driving modes excite more compression than others, particular regions in a cloud, and hence, also clouds \emph{on average}, can show characteristics of the flows produced with ideal driving with different mixtures of solenoidal and compressive modes.


Finally, we comment on the relation between the DGMFs and column density PDFs. The column density PDFs of nearby clouds are log-normal below $N(\mathrm{H}_2) \lesssim 3 \times 10^{21}$ cm$^{-2}$. In the range $N(\mathrm{H}_2) = 3-25 \times 10^{21}$ cm$^{-2}$, they are in agreement with either powerlaws or (wide) log-normals (KT13). It is not established if the PDFs above $N(\mathrm{H}_2) \gtrsim 3 \times 10^{21}$ cm$^{-2}$ are log-normals (KT13) or powerlaws \citep[][see Fig. \ref{fig:pdfs}]{sch13}. Importantly, it follows from Eq. \ref{eq:dgmf-pdf} that  \emph{a log-normal PDF yields an exponential DGMF and a powerlaw PDF yields a powerlaw DGMF.} The simulated DGMFs in the range $N(\mathrm{H}_2) \gtrsim 3-25 \times 10^{21}$ cm$^{-2}$ appear exponential at the early stages. Therefore, the column density PDFs at these stages are close to log-normals. When the simulations evolve, the DGMFs become closer to powerlaws (see FK13). This means that the underlying column density PDF transits from a log-normal to a powerlaw. 

\section{Conclusions} 
\label{sec:conclusions}

We have examined the relationship between the dense gas mass fraction (DGMF), star formation, and turbulence properties in molecular clouds by comparing DGMFs derived from isothermal, magneto-hydrodynamic, self-gravitating turbulence simulations to observed ones. Our conclusions are as follows. 	
\begin{enumerate}

   \item Simulations predict close-to exponential DGMFs for molecular clouds in the column density range of $N(\mathrm{H}_2) = 3-11 \times 10^{21}$ cm$^{-2}$. The DGMF slopes span the range $\alpha = [-0.41, -0.023]$, being clearly steeper at the early stages of the simulations compared to the stages when stars are forming ($\mathrm{SFE} \geq 1$\%). These predictions are accurate on a 70\% level up to $N(\mathrm{H}_2) \approx 25 \times 10^{21}$ cm$^{-2}$.
   
   \item The DGMF slopes depend strongly on the turbulence driving ($b$). They depend less, but  significantly, on the exact SFE. The dependence on the SFE is stronger in magnetized than non-magnetized cases. Generally, the effect of the magnetic field to the DGMF is small. Also ${\cal M}_\mathrm{s}$ has a negligible effect on the slopes in the examined column density range. The statistical variations are comparable to those arising from varying SFE. However, how compressive the turbulence is (i.e., parameter $b$) is the largest single factor in determining the slope of the DGMF in the simulations. 
   
   \item The observed DGMFs can be used to constrain the turbulence driving parameter $b$. The DGMFs of nearby clouds are only reproduced by simulations that are driven by relatively non-compressive forcing, i.e., $b = 1/3$ or 0.4. The fully compressive simulations ($b = 1$) over-estimate the DGMFs greatly. Massive IRDCs can show flatter DGMFs that are in agreement with more compressive driving. The spread of the observed DGMFs cannot be explained by different SFEs and statistical variations alone. Variations in the clouds' average compression level offer one explanation to account for the observed spread. 
   
\end{enumerate}


\begin{acknowledgements}
The work of JK was supported by the Deutsche Forschungsgemeinschaft priority program 1573 ("Physics of the Interstellar Medium"). C. F. acknowledges a Discovery Projects Fellowship from the Australian Research Council (grant DP110102191).	
\end{acknowledgements}



\pagebreak 

\appendix

\section{Numerical effects on the DGMFs}

\subsection{Effect of sink particles}
\label{app:sinks}

Sink particles \citep{fed10sinks} in the simulations accrete material into them after their creation, and hence, affect the density structure of their immediate surroundings in the simulation (and the DGMFs). 
In the following, we consider the effects of sink particles to the DGMFs. 

As described in FK12, the sink particles are created on a certain, resolution-dependent volume density and always have a radius of 2.5 pixels in the native resolution of the simulation. It follows that the sink particles have a resolution-dependent minimum density, which can further be converted into a minimum mean column density. Sink particles are created when a series of collapse criteria are fulfilled (see FK12), and when the local volume density exceeds
\begin{equation}
\rho_\mathrm{sink} = \frac{\pi c_\mathrm{s}^2}{4Gr_\mathrm{sink}^2},
\label{eq:rho_sink}
\end{equation}
where $c_\mathrm{s}$ is the isothermal speed of sound and $r_\mathrm{sink}$ the radius of the sink particle. It follows that the mean column density of a sink particle at the moment of its creation is
\begin{equation}
\overline{\Sigma}_\mathrm{sink} = \frac{\rho_\mathrm{sink} V_\mathrm{sink}}{\pi r_\mathrm{sink}^2} = \frac{4}{3}  \rho_\mathrm{sink}    r_\mathrm{sink}.
\label{eq:sigma_sink}
\end{equation}
The sink particle properties are listed in Table \ref{tab:sinks} for different physical resolutions. 

\begin{table*}
\begin{minipage}[t]{\textwidth}
\caption{Simulation properties (adapted from FK13)}             
\label{tab:sinks}      
\centering                         
\renewcommand{\footnoterule}{}  
\begin{tabular}{l l c c c c c c c c c c c c}   
\hline\hline                 
\#  	& name\footnote{Parentheses after the names refer to the different random seeds used in the simulations.}	&	${\cal M}_\mathrm{s}$	& $b$	& box size 	& box size		& $M$\footnote{Total mass in the simulation box.}	& $B_0$\footnote{Mean magnetic-field strength in $z$-direction of the computational domain.}	& $\alpha_\mathrm{vir, 0}$	&	$\alpha_\mathrm{vir}$ &$n(\mathrm{H}_2)_\mathrm{sink}$  	& $N(\mathrm{H}_2)_\mathrm{sink}$ & $M_\mathrm{sink}$	\\
&   &	& 		&	$[\mathrm{cells}]$	& $[\mathrm{pc}]$		&	[M$_\odot$]		& [$\mu$G]	&	&	&$[10^4 \ \mathrm{cm}^{-3}]$		& $[10^{21} \ \mathrm{cm}^{-2}]$ & [M$_\odot$] \\
\hline
6	&	GT256sM5	& 	 5 	& 1/3	&    256	&  2		& $3.9 \times 10^2$	& 0	& 1.0		& 8.0		& 28					& 22        	& 0.60  \\
7	&	GT256mM5	& 	 5 	& 0.4	&    256	&  2		& $3.9 \times 10^2$	& 0	& 0.98	& 5.4		& 28					& 22        	& 0.60  \\
8	&	GT256cM5	& 	 5 	& 1	&    256	&  2		& $3.9 \times 10^2$	& 0	& 0.82	& 1.5		& 28					& 22        	& 0.60  \\
\hline
10	&	GT256sM10	& 	10	&1/3	&    256 	&  8 		& $6.2 \times 10^3$	& 0	& 1.1		& 12.		& 1.7 				& 5.6 	& 2.4	\\    
11	&	GT512sM10	& 	10	&1/3	&    512 	&  8		& $6.2 \times 10^3$	& 0	& 1.1		& 12.		& 6.9 				& 11   	& 1.2\\
12	&	GT512mM10(s1)	& 	10	&0.4	&    512 	&  8		& $6.2 \times 10^3$ & 0	& 1.1		& 4.5		& 6.9 				& 11   	& 1.2\\
13	&	GT512mM10B1(s1)	& 	10	&0.4	&    512 	&  8		& $6.2 \times 10^3$ & 1	& 1.1		& 5.4		& 6.9 				& 11   	& 1.2\\
14	&	GT512mM10(s2)	& 	10	&0.4	&    512 	&  8		& $6.2 \times 10^3$ & 0	& 1.2		& 8.4		& 6.9 				& 11   	& 1.2\\
15	&	GT512mM10B1(s2)	& 	10	&0.4	&    512 	&  8		& $6.2 \times 10^3$ & 1	& 1.2		& 9.5		& 6.9 				& 11   	& 1.2\\
16	&	GT256mM10(s3)	& 	10	&0.4	&    256 	&  8 		& $6.2 \times 10^3$	& 0	& 1.0		& 5.9		& 1.7 				& 5.6 	& 2.4	\\    
17	&	GT512mM10(s3)	& 	10	&0.4	&    512 	&  8		& $6.2 \times 10^3$ & 0	& 1.0		& 5.9		& 6.9 				& 11   	& 1.2\\
18	&	GT512mM10B1(s3)	& 	10	&0.4	&    512 	&  8		& $6.2 \times 10^3$ & 1	& 0.97	& 6.4		& 6.9 				& 11   	& 1.2\\
19	&	GT256mM10B3(s3)	& 	10	&0.4	&    256 	&  8 		& $6.2 \times 10^3$	& 3	& 0.81	& 8.4		& 1.7 				& 5.6 	& 2.4	\\    
20	&	GT512mM10B3(s3)	& 	10	&0.4	&    512 	&  8		& $6.2 \times 10^3$ & 3	& 0.83	& 8.7		& 6.9 				& 11   	& 1.2\\
21	&	GT256mM10B10(s3)& 	10	&0.4	&    256 	&  8 		& $6.2 \times 10^3$	& 10	& 0.79	& 6.6		& 1.7 				& 5.6 	& 2.4	\\    
23	&	GT256cM10	& 	10	&1	&    256 	&  8 		& $6.2 \times 10^3$	& 0	& 0.85	& 1.1		& 1.7 				& 5.6 	& 2.4	\\    
24	&	GT512cM10	& 	10	&1	&    512 	&  8		& $6.2 \times 10^3$ & 0	& 0.87	& 1.1		& 6.9 				& 11   	& 1.2\\
\hline
25	&	GT256sM20	& 	20	&1/3	&    256 	&  32		& $9.9 \times 10^4$ & 0	& 1.0		& 11.1	& 0.11 				& 1.4 	& 9.6\\
26	&	GT256mM20	& 	20	&0.4	&    256 	&  32		& $9.9 \times 10^4$ & 0	& 1.1		& 4.5		& 0.11 				& 1.4 	& 9.6\\
27	&	GT256cM20	& 	20	&1	&    256 	&  32		& $9.9 \times 10^4$ & 0	& 1.0		& 0.60	& 0.11 				& 1.4 	& 9.6\\
\hline
28	&	GT256sM50	& 	50	&1/3	&    256 	&  200	& $3.9 \times 10^6$ & 0	& 1.1		& 12		& $2.8 \times 10^{-3}$	& 0.22 	& 60\\
29	&	GT512sM50	& 	50	&1/3	&   512 	&  200	& $3.9 \times 10^6$ & 0	& 1.1		& 13		& $1.1 \times 10^{-2}$	& 0.44	& 30	\\ 
30	&	GT256mM50	& 	50	& 0.4	&    256 	&  200	& $3.9 \times 10^6$ & 0	& 1.0		& 7.0		& $2.8 \times 10^{-3}$	& 0.22 	& 60\\
31	&	GT512mM50	& 	50	& 0.4	&   512 	&  200	& $3.9 \times 10^6$ & 0	& 1.1		& 7.4		& $1.1 \times 10^{-2}$	& 0.44	& 30	\\ 
32	&	GT256cM50	& 	50	& 1	&    256 	&  200	& $3.9 \times 10^6$ & 0	& 0.95	& 0.54	& $2.8 \times 10^{-3}$	& 0.22 	& 60\\
33	&	GT512cM50	& 	50	& 1	&   512 	&  200	& $3.9 \times 10^6$ & 0	& 0.99	& 0.56	& $1.1 \times 10^{-2}$	& 0.44	& 30	\\ 
\hline                              
\end{tabular}
\end{minipage}
\end{table*}

The sink particle column densities listed in Table \ref{tab:sinks} represent levels below which the DGMFs are \emph{not affected by sink particles, regardless of whether the sinks are removed or not}. In the most conservative interpretation, the DGMFs are reliable only below these column density limits. Therefore, we use the upper limit of $N(\mathrm{H}_\mathrm{2}) = 11 \times 10^{21}$ cm$^{-2}$, which is the sink particle column density for the ${\cal M}_\mathrm{s} = 10$ simulations $512^3$ cells in size, in the analysis performed in this paper.

However, it is not at all certain that the DGMF shape immediately above $N(\mathrm{H}_2)_\mathrm{sink}$ is greatly affected by the sink particles. Above $N(\mathrm{H}_2)_\mathrm{sink}$, there are lines-of-sight whose \emph{column} density is higher than the sink particle column density, but the local \emph{volume} densities do not reach high enough values for sink particles to form. In fact, these lines-of-sight are greatly more numerous in the simulations compared to those that contain sinks, especially at early times when the overall SFE is low.

We dealt with sink particles in this work by disregarding the lines-of-sight affected by them directly from the simulation data. While this procedure, in principle, eliminates the effects of sink particles, it removes mass preferentially from high column densities, and hence potentially biases the DGMF \emph{downwards} (steepens it). Consequently, it is important to note that the \emph{flattening} of the DGMFs seen in the simulations (see Section \ref{subsec:sim-DGMFs}) at around $N(\mathrm{H}_2) \approx 10-15 \times 10^{21}$ cm$^{-2}$ cannot be due to sink particle treatment; any associated incompleteness would bias the determination downwards, not upwards.

\begin{figure}
   \centering
\includegraphics[bb = 0 0 500 350, clip=true, width=\columnwidth]{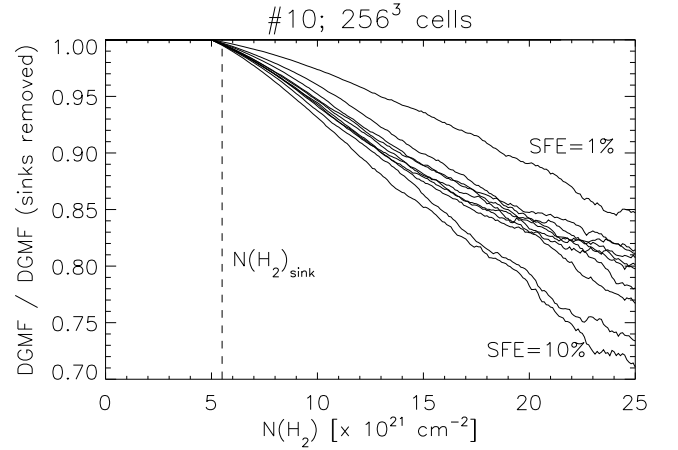}
      \caption{Error (incompleteness) in the derived DGMFs due to removal of sink particles. The figure shows the ratio of DGMFs derived with and without sink particle removal as a function of column density for time steps up to $\mathrm{SFE}=10\%$. The curves for $t=0$ and $\mathrm{SFE}=0\%$ are indistinguishable from unity. The plot is shown for simulation \#10 (${\cal M}_\mathrm{s}=10$, $256^3$ cells in size, $b = 1/3$). The error in other ${\cal M}_\mathrm{s}=10$ models is expected to be smaller, because of the higher sink particle column density and more compressive turbulence driving.  
                    }
         \label{fig:sinks}
   \end{figure}

We can quantify the incompleteness due to sink particle removal by comparing DGMFs derived with and without the elimination of sink particles. This experiment is shown in Fig. \ref{fig:sinks}, which shows the ratio of the DGMFs with and without the sink particle elimination as a function of column density. The plot is shown for the model in which the effect of sinks in the examined column density range is expected to be strongest, i.e., the solenoidal simulation with $256^3$ cell resolution. Higher resolution increases the sink particle column density (cf., Table \ref{tab:sinks}), and more compressive forcing increases the relative amount of high column densities, thereby reducing the error \emph{in the examined column density regime}. The figure shows that the error due to incompleteness (i.e., preferential removal of high-column densities) is less than 30\% below $N(\mathrm{H}_2) \lesssim 25 \times 10^{21}$ cm$^{-2}$ for SFEs up to 10\%. 

In summary, it can be concluded that the DGMFs derived for ${\cal M}_\mathrm{s}=10$ simulations are unaffected by the sink particles (or by their removal) below the $N(\mathrm{H}_2)_\mathrm{sink}$ values. In addition, the error in the predicted DGMFs is less than 30\% when the range up to $N(\mathrm{H}_2) \approx 25 \times 10^{21}$ cm$^{-2}$ is considered. 

\subsection{Effect of the simulation resolution}
\label{app:resolution}

The simulations of FK12 are either 128$^3$, 256$^3$, 512$^3$, or $1024^3$ computational cells in size. In this work, we used all but those simulations that are $128^3$ cells in size. It is possible that the different computational resolutions used in the simulations affect the DGMFs, as especially high column densities are potentially better resolved by higer-resolution simulations. We examined the possible effect of the simulation resolution to the DGMFs by comparing the DGMFs of simulations that were run with the same physical parameters, but have different computational resolution.

\begin{figure}
   \centering
\includegraphics[width=\columnwidth]{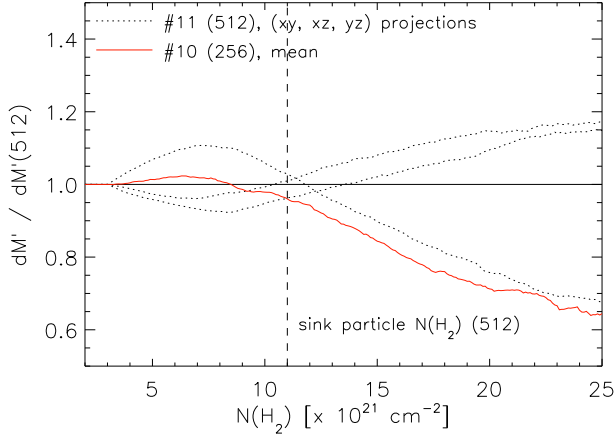}
      \caption{Effect of simulation resolution to the DGMFs. The red line shows the DGMF of simulation \#10 ($256^3$ cells in size) divided by the DGMF of simulation \#11 ($512^3$ cells in size). The physical parameters of the two simulations are the same. The dashed lines show the DGMFs calculated for different projections of model \#11 divided by the mean DGMF of model \#11.
                    }
         \label{fig:resotest}
   \end{figure}

Figure \ref{fig:resotest} shows as an example a comparison of DGMFs derived for models \#10 and \#11 that are $256^3$ and $512^3$ cells in size, respectively. All other parameters are same in these two models. The figure shows the DGMF of the model \#10 divided by that of model \#11 (red line). The figure also shows the DGMFs calculated for model \#11 using different projections (projections to xy, xz, and yz planes, black dotted lines). The DGMF of model \#10 is in good agreement with that of model \#11 below the sink particle column density, $N(\mathrm{H}_2) = 11 \times 10^{21}$ cm$^{-2}$. At higher column densities, the lower-resolution simulation (\#10) begins to under-estimate the column densities slightly. However, it is still within 30\% of the higher-resolution one up to the column density of $N(\mathrm{H}_2) \approx 25 \times 10^{21}$ cm$^{-2}$. We conclude that the effect of resolution is smaller than the uncertainty due to the projection effects in the column density range $N(\mathrm{H}_2) = 11 \times 10^{21}$ cm$^{-2}$ and accurate to 70\% level up to $N(\mathrm{H}_2) = 25 \times 10^{21}$ cm$^{-2}$.

\section{Illustration of column density PDFs}
\label{app:pdfs}

\begin{figure}
   \centering
\includegraphics[width=\columnwidth]{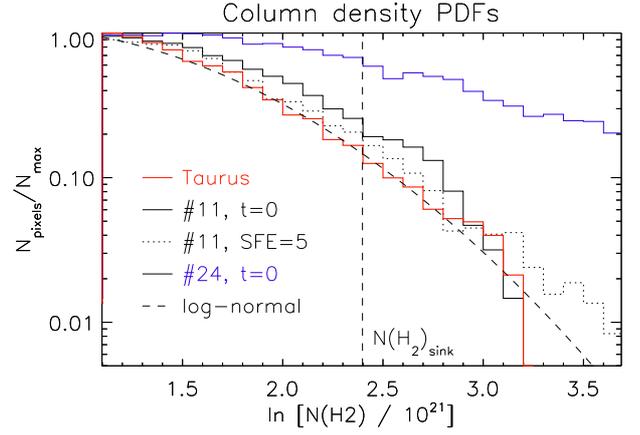}
      \caption{Column density PDFs of models \#11 ($b=1/3$) and \#24 ($b=1$), and the PDF of the Taurus molecular cloud. Both models have ${\cal M}_\mathrm{s} = 10$ and $B = 0$ $\mu$G, and they are $512^3$ computational cells in size. The black histograms show the PDFs of model \#11 at $t=0$ (solid line) and $\mathrm{SFE}=5\%$ (dotted line). The blue line shows the PDF of model \#24. The red line shows the PDF of Taurus from \citet{kai09}. Note that the dynamic range of the Taurus PDF ends at about $\ln{N(\mathrm{H}_\mathrm{2}) = 3.2}$. The black dashed line shows, for reference, a log-normal function. The PDFs in the range $N(\mathrm{H}_\mathrm{2}) = 3-11 \times 10^{21}$ cm$^{-2}$ can be described by a log-normal function, but also reasonably well by a powerlaw function (which would be a linear curve in the given presentation).  
                    }
         \label{fig:pdfs}
   \end{figure}

Figure \ref{fig:pdfs} show a comparison of the column density PDFs derived for models \#11 and \#24, and the PDF of the Taurus molecular cloud from \citet{kai09}. Note how the higher relative amount of high-column density material predicted by fully compressive simulations (\#24) is evidenced by a flatter PDF. In the column density range $N(\mathrm{H}_\mathrm{2}) = 3-25 \times 10^{21}$ cm$^{-2}$, the PDF of simulation \#11 is close to what is observed in Taurus. In this narrow range, the PDF is in a reasonable agreement with either a log-normal function (shown for a reference in the figure) or a powerlaw function. 

\end{document}